\documentclass{article}
\usepackage{spconf,amsmath,amssymb,graphicx,booktabs,multirow}
\usepackage{tikz}
\usetikzlibrary{arrows.meta,positioning,calc,fit}
\usepackage[hidelinks]{hyperref}
\usepackage{xcolor}
\usepackage{siunitx}

\newcommand{\ci}[2]{{\scriptsize\color{black!55}[#1,#2]}}

\title{VOCAL INTENSITY CONVERSION WITH CONTINUOUS SPL CONTROL \\ VIA ADVERSARIAL TRAINING}

\name{Quentin Le Tellier$^1$, Albert Rilliard$^1$, Olivier Perrotin$^2$, Marc Evrard$^1$%
      }
\address{$^1$Universit\'e Paris-Saclay, CNRS, LISN, Orsay, France\\
         $^2$Univ. Grenoble Alpes, CNRS, Grenoble INP, GIPSA-lab, Grenoble, France}

\begin{document}
\ninept
\maketitle

\begin{abstract}
Vocal intensity conversion renders a recorded utterance at a different vocal intensity, rather than simply scaling its amplitude, while preserving its content and speaker identity.
Existing systems control intensity through a reference utterance or a few discrete categories.
Instead, we control the conversion with any target sound pressure level (SPL) on a continuous dB scale.
A lightweight converter, trained as a conditional cycle-consistent GAN without parallel data, transforms the features of a frozen speech encoder, from which a pretrained vocoder resynthesizes the speech.
We compare three representations, log-mel spectrograms, WavLM, and SpeechTokenizer, and introduce an objective protocol that measures the achieved intensity change against the target one on a calibrated corpus.
With WavLM, the converter achieves a measured 83\% of the target change in intensity and also shifts the fundamental frequency.
A perceptual evaluation confirms that perceived vocal intensity increases with the target level on amplitude-normalized stimuli.
\end{abstract}


\section{Introduction}
\label{sec:intro}

Speakers adapt their vocal intensity to context, for instance, with distance or noise \cite{traunmuller2000acoustic}.
Speech technologies are mostly trained on speech produced at conversational effort, and underperform on speech produced at other levels \cite{zelinka2012impact}.
Being able to render a recorded utterance at a target intensity could provide targeted data augmentation and add a missing control dimension for expressive synthesis and cognitive evaluations.
We define the \textit{vocal intensity} as the sound pressure level (SPL) of a voice measured at one meter from the speaker's mouth \cite{lienard2019quantifying}, the principal measurable correlate of vocal effort.
It is an absolute acoustic quantity, so recovering it from a recording requires a calibrated chain and a known microphone distance \cite{svec2018tutorial}.
We propose a conversion model that renders a recorded utterance at a target vocal intensity while preserving its content and speaker identity.

Existing works approach the related but distinct problem of normal-to-Lombard conversion: mapping speech produced in quiet onto speech produced in
noise \cite{lombard1911signe}.
Early systems regress acoustic parameters and resynthesize through a vocoder \cite{perrotin16_interspeech}; later ones use adversarial style transfer \cite{seshadri2019cycle} or diffusion over self-supervised features \cite{zhu2024noise}.
All inherit the categorical framing of their corpora, restricting targets to discrete noise or effort conditions.

Two systems define the current state of the art and are most closely related to our setting. 
LombardTokenizer \cite{jacquelin25_interspeech} operates within a neural speech codec. 
It retrains SpeechTokenizer \cite{zhang2024speechtokenizer} so that the second residual-vector-quantization (RVQ) layer is distilled from a dedicated vocal-effort encoder.
Conversion is then performed via latent swapping: the second layer of the source is replaced with an encoded target utterance that carries the desired effort, keeping all other layers unchanged.
Instead, PGD-N2L \cite{chen2025pgdn2l} disentangles linguistic content, speaker identity, and the Lombard effect, representing the last by a learnable lookup table over the discrete noise levels of its corpus.
Control at inference involves choosing a target noise level, whose embedding is injected into the content stream.

We depart from this literature in three ways.
First, we condition on a physical quantity: the equivalent continuous level of the utterance in dB SPL referred to 1\,m, rather than a reference recording that already carries the desired effort \cite{jacquelin25_interspeech} or a noise category \cite{chen2025pgdn2l}.
Being a scalar, it is well defined for any value within the training range and carries no property of a reference utterance into the output.
Second, we keep the encoder and the decoder frozen and train only a small converter that transforms the encoded frames, rather than retraining the representation.
Third, we measure whether a target level was \emph{achieved}, which neither system can do since it lacks a target in physical units.

A separate line of work makes a direct measurement of the intensity of an uncalibrated speech recording possible, combining calibrated corpora \cite{alku2024avid} with models that recover vocal intensity from amplitude-normalized speech \cite{letellier25_interspeech, kodali2025automatic}. 
We adopt such a predictor, frozen and absent from the generator objective, to report the achieved level of each conversion relative to the target.

Our contributions are (i) a formulation of vocal intensity conversion as conditioning on a continuous target in dB SPL; (ii) a conditional cycle-consistent generative adversarial network (GAN), trained without parallel data or an auxiliary intensity predictor; (iii) an evaluation of this conversion method across three frozen encoder/decoder pairs; and (iv) objective and perceptual evaluation protocols for vocal intensity conversion.
The code, model weights, and converted test samples are publicly available.\footnote{\url{https://github.com/vers-project/vocal-intensity-conversion}}

\section{Method}
\label{sec:method}

\subsection{Task formulation}
\label{ssec:task}

Given the recording $x$ of one utterance and a target vocal intensity level $\tau_t$, we aim to convert the utterance as the speaker would have produced it at $\tau_t$, preserving content and speaker identity. 
The target is a scalar in dB SPL drawn from a continuous range.
For that request to be well posed, the level must be a property of the voice rather than of the recording setup. 
Two conditions on the recording are sufficient: the recording chain is calibrated, so that a digital sample can be converted into an absolute sound pressure; and the mouth-to-microphone distance is known, so that pressure can be referred to a distance of \SI{1}{\meter}, at which levels measured in different sessions and across different speakers become comparable.
The vocal intensity of $x$ is then its equivalent continuous level ($L_{eq}$, \cite{svec2018tutorial}):
\begin{equation}
  \tau(x) \;=\; 10 \log_{10}\!\left(\frac{1}{N}\sum_{n=1}^{N}
                 \left(\frac{p_n}{p_0}\right)^{\!2}\right)
  \quad [\mathrm{dB\ SPL}],
  \label{eq:leq}
\end{equation}
with $p_n$ the root-mean-square (RMS) pressure over the $n$-th of $N$ \SI{25}{\milli\second} windows and $p_0 = \SI{20}{\micro\pascal}$ the reference pressure.
We write $\tau_s = \tau(x)$.

\subsection{Conversion in a frozen feature space}
\label{ssec:frozen}
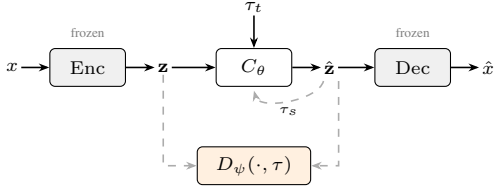
\begin{figure}[t]
  \centering
  \begin{tikzpicture}[
      x=1mm, y=1mm,
      box/.style   = {draw, rounded corners=2pt, minimum width=10mm,
                      minimum height=5mm, align=center, font=\scriptsize},
      frozen/.style= {box, fill=gray!12},
      critic/.style= {draw, rounded corners=2pt, minimum width=15mm,
                      minimum height=5mm, align=center, font=\scriptsize,
                      fill=orange!12},
      sym/.style   = {font=\scriptsize, inner sep=1pt},
      arr/.style   = {-{Stealth[scale=0.7]}, semithick},
      darr/.style  = {-{Stealth[scale=0.7]}, semithick, dashed, gray!65},
    ]
    \node[sym]    (x)    at (0,0)    {$x$};
    \node[frozen] (enc)  at (10,0)   {$\mathrm{Enc}$};
    \node[sym]    (z)    at (20,0)   {$\mathbf{z}$};
    \node[box]    (conv) at (32,0)   {$C_\theta$};
    \node[sym]    (zh)   at (42,0)   {$\hat{\mathbf{z}}$};
    \node[frozen] (dec)  at (53,0)   {$\mathrm{Dec}$};
    \node[sym]    (xh)   at (63,0)   {$\hat{x}$};

    \draw[arr] (x) -- (enc);   \draw[arr] (enc) -- (z);
    \draw[arr] (z) -- (conv);  \draw[arr] (conv) -- (zh);
    \draw[arr] (zh) -- (dec);  \draw[arr] (dec) -- (xh);

    \node[font=\tiny, gray, above=0.3mm of enc] {frozen};
    \node[font=\tiny, gray, above=0.3mm of dec] {frozen};

    \node[sym] (taut) at (32,8) {$\tau_t$};
    \draw[arr] (taut) -- (conv);

    \draw[darr] ($(zh.south)+(-0.8,-0.5)$) to[out=-105, in=-75, looseness=0.75]
        node[midway, below, font=\tiny, black, inner sep=0.7mm] {$\tau_s$}
        ([yshift=-0.5mm]conv.south);

    \node[critic] (d) at (32,-13) {$D_\psi(\cdot,\tau)$};
    \coordinate (zdown)  at (20,-13);
    \coordinate (zhdown) at (43.2,-13);
    \draw[darr] (z.south) -- (zdown) -- (d.west);
    \draw[darr] ($(zh.south)+(1.2,-0.5)$) -- (zhdown) -- (d.east);

  \end{tikzpicture}
  \caption{Modeling strategy: $\mathrm{Enc}$ and $\mathrm{Dec}$ are frozen; $C_\theta$ is trained. The dashed arc is the cycle: the converted frames pass back through the same converter at the source level and must return $\mathbf{z}$. $D_\psi$ scores (sequence, level) pairs (Sec.~\ref{ssec:cgan}). At inference, only the top row runs.}
  \label{fig:system}
\end{figure}

We convert speech in the feature space of a frozen encoder/decoder pair rather than in the waveform domain.
An encoder $\mathrm{Enc}$ analyzes a speech signal into a sequence of frames
$\mathbf{z} = (z_1, \dots, z_T)$, $z_i \in \mathbb{R}^{d}$, and a decoder
$\mathrm{Dec}$ synthesizes a signal from such a sequence, so that
$\mathrm{Dec}(\mathrm{Enc}(x)) \approx x$. A converter $C_\theta$
transforms this sequence. The conversion of $x$ to $\tau_t$ is then
\begin{equation}
  \hat{x} \;=\; \mathrm{Dec}\bigl(C_\theta(\mathrm{Enc}(x),\, \tau_t)\bigr),
  \label{eq:pipeline}
\end{equation}
where $\mathrm{Enc}$ and $\mathrm{Dec}$ are frozen throughout (Fig.~\ref{fig:system}) and only $C_\theta$ is trained. 
The output, therefore, inherits the reconstruction quality of the chosen encoder/decoder pair.

$C_\theta$ is frame-synchronous: it emits one output frame per input frame, so
conversion preserves the alignment of the sequence and cannot alter duration or
speaking rate.
$C_\theta$ is also residual: $C_\theta(\mathbf{z}, \tau_t) = \mathbf{z} + \Delta_\theta(\mathbf{z}, \tau_t)$, where $\Delta_\theta$ predicts the correction from the whole sequence.
The formulation also admits a vector $\tau_t$, one value per frame, which would give control over the intensity contour; here $\tau_t$ is a single value per utterance.

\subsection{$\tau$-conditional training objective}
\label{ssec:cgan}

$C_\theta$ is trained as the generator of a conditional GAN. Let $p(\mathbf{z}, \tau_{s})$ denote the joint distribution of a real frame sequence and the source level at which it was produced, and $q(\hat{\mathbf{z}}, \tau_{t})$ the corresponding distribution of converted sequences and their target levels. A discriminator $D_\psi(\mathbf{z}, \tau)$ is trained to separate samples of $p$ from a mixture of samples of $q$ and of mismatched real pairs, in which a real sequence is paired with a vocal intensity level other than its own.

\textbf{Conditional discriminator.} Miyato and Koyama~\cite{miyato2018cgans} argue that when a generator learns the condition from the loss of an auxiliary classifier, whether trained jointly with the discriminator or pretrained, it is encouraged to produce samples that the classifier recognizes easily, and its conditional distribution moves away from that of the data.
Using a pretrained intensity predictor as the training signal for the level, either alone or alongside an unconditional discriminator, has the same structure: the relation between the sequence and the level is then set by the predictor rather than by real data.
We therefore adopt their alternative and condition $D_\psi$ itself on $\tau$, so that it targets the joint distribution $p(\mathbf{z}, \tau)$.

Training alternates between a discriminator step and a generator step, both using the least-squares GAN loss \cite{mao2017lsgan}, which regresses the score toward $+1$ for a pair $D_\psi$ should accept and toward $-1$ for one it should reject.
Each discriminator step scores three pairs: a positive pair (a real sequence with its source level $\tau_+$), and two negative pairs (a converted sequence with its target level $\tau_t$, and a real sequence with a mismatched level $\tau_-$):
\begin{equation}
  \begin{split}
  \mathcal{L}_D \;=\; &\tfrac{1}{2}\,
      \mathbb{E}\bigl[(D_\psi(\mathbf{z}, \tau_+) - 1)^2\bigr]
   \;+\; \tfrac{1}{4}\,
      \mathbb{E}\bigl[(D_\psi(\hat{\mathbf{z}}, \tau_t) + 1)^2\bigr] \\
   &+\; \tfrac{1}{4}\,
      \mathbb{E}\bigl[(D_\psi(\mathbf{z}, \tau_-) + 1)^2\bigr],
  \end{split}
  \label{eq:dloss}
\end{equation}
where the three weights give the positive and the two negatives equal total mass.
Spectral normalization \cite{miyato2018spectral} is applied to $D_\psi$.
Both real pairs are built from $\tau_s$, so training requires the level at which
every training utterance was produced, which we measure from the calibrated
recording by Eq.~\eqref{eq:leq}.

\textbf{Discriminator training pairs.}
As in any conditional GAN, $D_\psi$ learns which levels are compatible with a real sequence.
Because $\tau$ is continuous, each measured level occurs only once in the training data, and the target levels given to $C_\theta$ generally match no training sequence.
Following CcGAN~\cite{ding2021ccgan}, we therefore treat a real sequence as a valid example of every level within $v$ of its source level: the positive pair uses $\tau_+ = \tau_s + \mathcal{U}(-v, v)$.
Mismatched levels $\tau_-$ are drawn at least $\delta > v$ away from $\tau_s$, so that no pairing can be both a positive and a negative.

\textbf{Content preservation.} 
Cycle consistency requires that converted frames, passed back through the generator at the source level, recover $\mathbf{z}$: if the generator altered the content or the speaker identity, the return path cannot restore them \cite{zhu2017cyclegan}.
An identity anchor requires the generator to return $\mathbf{z}$ unchanged when the target level matches the level already present. 
The generator minimizes
\begin{equation}
  \begin{aligned}
  \mathcal{L}_G \;=\;\; & \mathbb{E}\bigl[(D_\psi(\hat{\mathbf{z}}, \tau_t) - 1)^2\bigr]
      && \text{(adversarial)} \\
   &+ \mathbb{E}\bigl[\lVert C_\theta(\hat{\mathbf{z}}, \tau_s) - \mathbf{z}\rVert^2\bigr]
      && \text{(cycle)} \\
   &+ \mathbb{E}\bigl[\lVert C_\theta(\mathbf{z}, \tau_s) - \mathbf{z}\rVert^2\bigr].
      && \text{(identity)}
  \end{aligned}
  \label{eq:loss}
\end{equation}

\section{Experimental setup}
\label{sec:setup}

\subsection{Corpus}
\label{ssec:corpus}

We use AVID \cite{alku2024avid}, a calibrated database of speech produced at varying vocal intensities, in which fifty speakers (25 female, 25 male) read 25 sentences and two paragraphs at four levels (soft, normal, loud, and very loud) spanning roughly 40\,dB. 
Calibration makes Eq.~\eqref{eq:leq} absolute for recordings. 
The speakers followed instructions rather than a prescribed level: each category covers a speaker-dependent range of intensity. 
We segment AVID's continuous recording sessions (17.94\,h across both tasks), so that the training data keeps its natural pauses. 
We split the speakers 40/4/6 into training, validation, and test sets, disjoint by speaker and stratified by sex.
Although AVID is parallel across levels, training uses no parallel data: the converter output is never explicitly trained to match another recording.

\subsection{Encoder/decoder pairs}
\label{ssec:pairs}

To determine how the feature space affects the conversion, and whether some spaces are better suited to the task than others, we train the same converter architecture, with the same objective and corpus, on three encoder/decoder pairs and compare the results. The first analyzes speech into an 80-band log-mel spectrogram and inverts it with BigVGAN-v2 \cite{lee2023bigvgan}, in its
22.05\,kHz, 80-band, 8\,kHz-bandwidth configuration. 
The second takes the sixth layer of WavLM-Large \cite{chen2022wavlm} and inverts it with the HiFi-GAN \cite{kong2020hifigan} that kNN-VC \cite{baas23_interspeech} provides. 
The third is SpeechTokenizer \cite{zhang2024speechtokenizer}, used at the continuous output of its encoder, before quantization. 
The log-mel encoder runs at 86 frames per second with $d = 80$; the other two operate at 16\,kHz and produce 50 frames per second with $d = 1024$. All three are used at their published weights, without fine-tuning.\footnote{\url{https://huggingface.co/nvidia/bigvgan_v2_22khz_80band_fmax8k_256x}; regular (not prematched) HiFi-GAN from \url{https://github.com/bshall/knn-vc}; HuBERT-avg checkpoint from \url{https://huggingface.co/fnlp/SpeechTokenizer}.}

The three feature spaces differ in nature: the log-mel spectrogram is an acoustic representation, typically used in voice conversion; SpeechTokenizer's is the latent space of a neural codec; and WavLM layer 6 is a self-supervised representation closely related to phonetic content, which kNN-VC exploits for voice conversion.

\subsection{Training}
\label{ssec:training}

$C_\theta$ and $D_\psi$ are six-layer bidirectional
Transformer encoders \cite{vaswani2017attention} of model width 128 with four attention heads, of dimension 128 in $C_\theta$ and 32 in $D_\psi$, and rotary position embeddings \cite{su2024roformer}. 
$D_\psi$ pools its output by a mean over frames and is conditioned on $\tau$ by projection \cite{miyato2018cgans}. 
In both networks,
$\tau$ is standardized, encoded at 64 logarithmically spaced frequencies, and passed through a two-layer MLP. 
$D_\psi$ uses L2 rather than dot-product attention.

Training runs on 2\,s excerpts, with $\tau_t$ drawn uniformly from $[42.8, 76.1]$\,dB, the 5th to 99th percentile of $\tau$ over the training excerpts. 
The lower tail is trimmed harder because about 3\% of excerpts fall entirely inside a pause. 
We set $v = 2.5$\,dB and $\delta = 6$\,dB. 
Both networks use Adam with $\beta_1 = 0$ and $\beta_2 = 0.99$, at constant learning rates of $2\!\cdot\!10^{-4}$ for $C_\theta$ and $5\!\cdot\!10^{-5}$ for $D_\psi$, batches of 128 excerpts per GPU, and gradients clipped at 1.0. 
Each converter is trained for 11\,400 iterations on four H100 GPUs using distributed data parallelism, taking about 3.7\,h for WavLM~L6 and SpeechTokenizer and about 9\,h for log-mel.

\subsection{The intensity predictor}
\label{ssec:predictor}

The intensity predictor $P_\phi$ is a Transformer regression head (six layers, width 128, four heads) over the second layer of a frozen wav2vec~2.0 encoder \cite{baevski2020wav2vec}.
Its input is normalized to zero mean and unit variance per utterance, so amplitude is not available as a cue.
Its design follows \cite{letellier25_interspeech}, where wav2vec~2.0 embeddings were shown to support vocal intensity regression that transfers across calibrated corpora in different languages. 
It was trained on the same 40 speakers as the converter, so the six test speakers are unseen by both. 
On those six speakers, it estimates the level of an utterance with an RMSE of 1.43\,dB and a mean error of $+0.47$\,dB.

\begin{table*}[t]
  \caption{Displacement slopes per encoder/decoder pair; brackets give 95\% confidence intervals over speakers. WER and UTMOS are absolute values, not slopes (0.047 and 3.65 on the original recordings). Bold marks the best \emph{converted} entry per column (slope closest to one) and any tied with it under a speaker-paired test.}
  \label{tab:slopes}
  \centering\footnotesize\setlength{\tabcolsep}{2pt}
  \begin{tabular}{@{}ll ccccc cc@{}}
    \toprule
    Encoder/decoder pair & & intensity ($P_\phi$) & RMS & $f_0$ & F1 & alpha ratio & WER & UTMOS \\
    \midrule
    \multirow{2}{*}{Log-mel}
      & converted & 0.62\,\ci{0.58}{0.66} & \textbf{1.08}\,\ci{1.00}{1.18} & \textbf{0.42}\,\ci{0.28}{0.62} & \textbf{0.65}\,\ci{0.59}{0.72} & \textbf{0.75}\,\ci{0.66}{0.85} & 0.136\,\ci{0.095}{0.176} & 2.60\,\ci{2.37}{2.86} \\
      & resynthesis &0.97\,\ci{0.96}{0.99} & 1.00\,\ci{1.00}{1.00} & 0.98\,\ci{0.94}{1.00} & 0.98\,\ci{0.97}{0.99} & 0.99\,\ci{0.99}{1.00} & 0.047\,\ci{0.033}{0.058} & 3.56\,\ci{3.40}{3.74} \\
    \addlinespace
    \multirow{2}{*}{WavLM L6}
      & converted & \textbf{0.83}\,\ci{0.78}{0.89} & \textbf{1.04}\,\ci{0.98}{1.09} & \textbf{0.48}\,\ci{0.32}{0.68} & 0.43\,\ci{0.37}{0.51} & 0.43\,\ci{0.37}{0.47} & \textbf{0.083}\,\ci{0.059}{0.109} & 2.89\,\ci{2.60}{3.17} \\
      & resynthesis &1.01\,\ci{0.90}{1.11} & 1.03\,\ci{0.98}{1.08} & 0.92\,\ci{0.80}{1.00} & 0.95\,\ci{0.87}{1.03} & 0.89\,\ci{0.85}{0.94} & 0.067\,\ci{0.047}{0.088} & 3.14\,\ci{2.82}{3.42} \\
    \addlinespace
    \multirow{2}{*}{SpeechTokenizer}
      & converted & 0.58\,\ci{0.51}{0.66} & 0.73\,\ci{0.67}{0.81} & 0.13\,\ci{0.07}{0.22} & 0.42\,\ci{0.36}{0.48} & 0.63\,\ci{0.57}{0.71} & 0.155\,\ci{0.108}{0.195} & \textbf{3.07}\,\ci{2.81}{3.32} \\
      & resynthesis &0.90\,\ci{0.85}{0.94} & 0.95\,\ci{0.90}{0.98} & 0.94\,\ci{0.86}{0.99} & 0.89\,\ci{0.84}{0.96} & 0.97\,\ci{0.95}{1.00} & 0.124\,\ci{0.083}{0.163} & 3.14\,\ci{2.88}{3.38} \\
    \bottomrule
  \end{tabular}
\end{table*}

\subsection{Evaluation protocol}
\label{ssec:protocol}

In the objective evaluation, each recording is converted to the measured level of another recording of the same speaker and sentence, and the conversion is compared with that recording.
As AVID's sentence list was read twice, a group is one speaker's four readings of one sentence within a single repetition.
The models convert each recording in a group to each of the four measured levels of that group, including its source level, yielding 16 conversions per group.
Over the six test speakers, 25 sentences, and two repetitions, this yields 300 groups, 1198 source recordings, and 4786 conversions. 
Three conditions are evaluated: the original recordings, the result of passing them through the encoder/decoder pair without conversion (\emph{resynthesis}), and the conversions.

For a measure $m$ and two original recordings $x_s$ and $x_t$ of a group, let $\tilde{x} = \mathrm{Dec}(\mathrm{Enc}(x))$ denote a resynthesis and $\hat{x}$ the conversion of $x_s$ to $\tau_t = \tau(x_t)$. 
Within each speaker, we perform a linear regression of the achieved displacement $m(\hat{x}) - m(\tilde{x}_s)$ on the target displacement $m(x_t) - m(x_s)$.
A slope of 1 means the target change was fully realized. 
The intercept reveals a systematic offset, and the residual spread informs on consistency.
The same fit applied to the resynthesis alone, $m(\tilde{x}_t) - m(\tilde{x}_s)$, indicates how much of the target change the representation carries before any conversion is learned. 
We call this slope the ceiling.
Confidence intervals are bootstrapped over the six speakers.

Displacement is measured primarily on the level predicted by $P_\phi$, but also on other acoustic parameters, to assess whether the model induces the acoustic changes expected in natural speech (e.g., \cite{rilliard2018paradigmatic}). 
It was also measured on RMS amplitude, $f_0$, and the first formant ($F_1$, estimated with Praat \cite{boersma2024praat}), as well as the $\alpha$-ratio, a spectral slope estimation from eGeMAPS \cite{eyben2016geneva}.
$f_0$ and $F_1$ are expressed in semitones.

Preservation is reported separately: word error rate (WER) from Whisper large-v3 \cite{radford2023whisper}, predicted naturalness from UTMOS \cite{saeki22c_interspeech}, and speaker similarity from ECAPA-TDNN \cite{desplanques20_interspeech} are estimated against a reference set by the two original recordings at those same two levels, since a change in vocal intensity displaces a speaker embedding even in natural speech.

\subsection{Perceptual evaluation protocol}
\label{ssec:perceptual}

We assess the change in perceived vocal intensity produced by the conversions through a perceptual experiment.
We selected the two female and two male test speakers with the widest vocal intensity range.
Two sentences were taken from each, in different vocal intensity categories, so that the eight source recordings cover eight sentences and each category twice.
Each source recording was converted to eight evenly spaced target levels over the range 45--75\,dB SPL, yielding 64 stimuli.
On each trial, a listener heard a converted stimulus and an original recording of the same sentence by the same speaker, and chose the one perceived as louder.
All stimuli were peak-normalized so that listeners judge vocal intensity from voice quality rather than from playback amplitude.
The original recordings, called anchors, are the four readings of that sentence in its second repetition, excluding the source recording.
Pairing each converted stimulus with the four anchors yields 256 pairs.
Thirty-two control pairs compare each source recording with the same four anchors.
The 288 pairs were divided into four disjoint sets of 72 by a Latin square, and each listener heard one set.
Eighty-one listeners were recruited on the Prolific platform.
In each set, five of the eight control pairs differ by at least 6\,dB in measured level, and the 54 listeners who answered at least four of these five correctly were retained.

The measured levels of the anchors form a scale on which each conversion is placed.
Let $\mu$ denote the perceived level of a converted stimulus and $\ell$ the measured level of an anchor.
The probability of judging the conversion louder is modeled as $\Phi\!\left(\beta(\mu-\ell)\right)$, with $\Phi$ the standard normal distribution function and $1/\beta$ the spread of the comparisons in decibels.
Probit regression estimates $\mu$, the point of subjective equality (the anchor level at which both members are equally likely to be chosen \cite{Wichmann2001}), which we report as the perceived level in dB SPL.
Confidence intervals are computed by resampling listeners instead of trials, since all listeners in a set judge the same 72 pairs.

\section{Results}
\label{sec:results}

Table~\ref{tab:slopes} reports the displacement slopes of each encoder/decoder pair.
On the level predicted by $P_\phi$, WavLM L6 reaches 0.83, against 0.62 for log-mel and 0.58 for SpeechTokenizer, and the ranking is unchanged relative to each representation's ceiling.
The WavLM L6 converter thus performs best on the estimated vocal intensity displacement; as vocal intensity is the property the task targets, its conversions are the ones we evaluate perceptually.

Beyond the estimated level, the other measures show which speech attributes the conversion changes.
All three converters displace the RMS amplitude, SpeechTokenizer to a lesser extent (0.73).
Amplitude alone, however, could be changed by a broadband gain without altering the voice.
The fundamental frequency, which such a gain leaves unchanged, is displaced by WavLM L6 (0.48) and log-mel (0.42), but hardly by SpeechTokenizer (0.13, against a ceiling of 0.94).
The spectral measures favor log-mel instead, on both the alpha ratio (0.75) and F1 (0.65).
Every representation preserves these attributes in resynthesis, with slopes of at least 0.89, so the information is not lost in the encoded features; the gaps between converters instead indicate that an attribute is more controllable in some feature spaces than in others.
No feature space therefore leads on every measure.

\textbf{Preservation.} 
For every representation, conversion increases the WER and lowers predicted naturalness relative to resynthesis.
WavLM L6 has the lowest WER after conversion, and SpeechTokenizer has the highest UTMOS; log-mel is nearly transparent in resynthesis but degrades most under conversion.
Speaker similarity is 0.61 for WavLM L6 conversions, against 0.73 between the original recordings at the source and target levels, which differ only in vocal intensity; the conversions thus retain most of the speaker similarity found between original recordings.

\begin{figure}[t]
  \centering
  \includegraphics[width=0.6\linewidth]{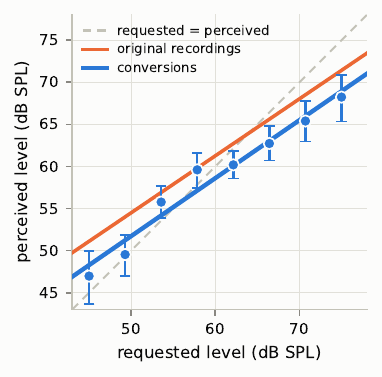}
  \caption{Perceived level of the conversions against their target level, both in dB SPL, from 54 listeners. Points are the fitted perceived levels, with intervals resampled across listeners, and the dashed diagonal marks a conversion heard at the target level. Original recordings measured on the same scale give the upper line.}
  \label{fig:perceived}
\end{figure}

\textbf{Perceptual evaluation.}
Figure~\ref{fig:perceived} shows the perceived level of the WavLM L6 conversions against the target level.
Perceived level increases with the target over the whole range, with a slope of 0.69 (95\% interval 0.60--0.78): the 30\,dB target range is perceived as a 21\,dB range.
Original recordings give a slope of 0.68 (0.51--0.84), which does not differ from that of the conversions, so listeners scale converted changes in vocal intensity as they scale natural ones.
Within each source recording, the eight conversions are also perceived in the target order, with rank correlations ranging from 0.62 to 0.95.
The absolute perceived level is less consistent: for the same target, it varies between source recordings with a standard deviation of about 6\,dB.
The conversion, therefore, controls relative vocal intensity well, whereas its absolute perceived level depends on the source recording.

\section{Conclusion}
\label{sec:conclusion}
We formulated vocal intensity conversion as conditioning on a continuous target in dB SPL.
This target is enforced adversarially by a conditional cycle-consistent GAN operating in the feature space of a frozen encoder/decoder pair.
Of the three feature spaces compared, WavLM L6 achieves 83\% of the target intensity change, and the perceived level of its conversions increases with the target level.
This work has three limitations.
First, the model controls vocal intensity relative to the source recording but is less consistent in reaching an absolute perceived level.
Second, the conversions displace $f_0$ and the spectral measures to a lesser extent than the original recordings do, which leaves room for improvement.
Third, calibrated corpora covering a wide range of vocal intensities are scarce, which limits the training data.

In future work, since no feature space leads on every measure, a hybrid feature set with a dedicated decoder could combine their strengths.
More importantly, estimating the training source levels with an automatic intensity predictor would address the third limitation.
Training could then use larger, uncalibrated corpora covering a wider range of vocal intensities, with an independent intensity predictor for evaluation.

\vfill\pagebreak

\section{Acknowledgments}
This work was partly funded by the French National Research Agency (ANR) through the VERS project (ANR-23-CE38-0010-01).
This work was granted access to the HPC resources of IDRIS under the allocation 2026-AD011016051R1 made by GENCI.

\bibliographystyle{IEEEbib_short}
\bibliography{refs}

\end{document}